\documentstyle[12pt]{article}
\begin{document}

\begin{center}
{Nearly Antiferromagnetic Fermi Liquids:  A Progress Report}\\

\vspace{1cm}

{David Pines}\\

{\it Center for Nonlinear Studies,\\ Los Alamos National Laboratory,
Los Alamos, NM  87545\\
and\\
Department of Physics,
 University of Illinois at Urbana-Champaign\\ 1110 West Green Street, Urbana, IL
61801-3080 USA}\\

\vspace{1cm}

\end{center}

\begin{abstract}
I describe recent theoretical and experimental progress in
understanding the physical properties of the two
dimensional nearly antiferromagnetic Fermi liquids (NAFL's) found in the normal
state of the cuprate superconductors.  In such NAFL's, the magnetic interaction
between planar quasiparticles is strong and peaked at or near the commensurate
wave vector, ${\bf Q} \equiv (\pi,\pi)$.  For the optimally doped and underdoped
systems, the resulting strong antiferromagnetic correlations produce three
distinct magnetic phases in the normal state:  mean field above $\rm T_{cr}$,
pseudoscaling between $\rm T_{cr}$ and $\rm T_*$, and pseudogap below $\rm
T_*$.  I present arguments which suggest that the physical origin of the
pseudogap found in the quasiparticle spectrum below $\rm T_{cr}$ is the
formation of a precursor to a spin-density-wave-state, describe the
calculations based on this scenario of the dynamical spin susceptibility, Fermi
surface evolution, transport, and Hall effect, and summarize the experimental
evidence in its support.
\end{abstract}

\def\alt{\mathrel {\mathpalette \vereq <}}
\def\agt{\mathrel {\mathpalette \vereq >}}
\def\mathpalette#1#2{\mathchoice {#1\displaystyle {#2}}%
                                 {#1\textstyle {#2}}%
                                 {#1\scriptstyle {#2}}%
                                 {#1\scriptscriptstyle
{#2}}}
\def\vereq#1#2{\lower 3pt\vbox {\baselineskip 1.5pt \lineskip 1.5pt \ialign
{$\m@th #1\hfill ##\hfil $\crcr #2\crcr \sim \crcr }}}
\def\m@th{\mathsurround=0pt}

\newpage

\section*{\bf Introduction}

A decade of experiments have shown us that nearly all the normal state
properties of the cuprate superconductors are anomalous when compared to the
Landau Fermi liquids found in the normal state of conventional superconductors.
 It is generally agreed that the mechanism for high temperature
superconductivity must be directly related to this unusual normal state
behavior.  The proposal that it is the magnetic interaction between planar
quasiparticles which is responsible for their anomalous normal state behavior
and the transition at high temperatures to a superconducting state with $\rm
d_{x^2-y^2}$ pairing was made some seven years ago at a conference on strongly
correlated electron systems. \cite{1}  It was based on the first generation of
nuclear magnetic resonance (NMR) experiments, which showed that the
quasiparticles in even optimally doped systems such as YBa$_2$Cu$_3$O$_7$ and
La$_{1.85}$Sr$_{0.15}$ displayed almost antiferromagnetic behavior; it led to
the ansatz that the normal state is best described as a nearly
antiferromagnetic Fermi liquid (NAFL) in which the effective magnetic
interaction between planar quasiparticles mirrors the highly anisotropic
momentum dependence found in NMR measurements of the spin-spin response
function.  In this lecture, I review recent theoretical and experimental
progress
in characterizing the behavior of two-dimensional NAFL's [a detailed review of
work on the NAFL prior to 1995 may be found in Ref. (2)], and in understanding
the normal state of the optimally doped and underdoped cuprate superconductors.

\section*{\bf Magnetic Behavior and Phase Diagram}

Of the various anomalous aspects of normal state behavior of the
superconducting cuprates, the low frequency magnetic response is
perhaps the most unusual, in that one finds nearly antiferromagnetic behavior
and three distinct magnetic phases in all but the highly overdoped systems.  A
quantitative fit to NMR and INS experiments can be obtained with a
phenomenological expression for the dynamical spin susceptibility,
$\chi(\rm {\bf q},\omega)$, which reflects this  close approach to
antiferromagnetism.  Quite generally, one finds four peaks in $\chi$ at wave
vectors, $\rm {\bf Q}_i$, in the vicinity of the commensurate AF
wave vector, $\rm {\bf Q} = (\pi/a,\pi/a)$, which are located
symmetrically about $\rm {\bf Q}$. \cite{3}  In the vicinity of a given peak at
$\rm {\bf Q}_i$,
$\chi$ displays considerable structure; it takes the form proposed by Millis
{\it et al.} \cite{4},

\begin{equation}
\rm \chi_{NAFL}({\bf q},\omega) = {\chi_{Q_i} \over
1+({\bf Q}_i-{\bf q})^2\xi^2-i\omega/\omega_{SF}}
\end{equation}

\noindent where as one approaches the superconducting transition
temperature $\rm T_c$, $\rm \chi_{{\bf Q}_i} \equiv \alpha \xi^2$, the static
peak susceptibility, is some orders of magnitude larger than the uniform
susceptibility, $\rm \chi_0$ and $\xi$, the antiferromagnetic correlation
length, is large compared to the
lattice spacing, a.  For example, in La$_{1.86}$Sr$_{0.14}$CuO$_4$, just
above $\rm T_c$, neutron scattering eperiments \cite{5} show that $\chi_{{\bf
Q}_i} = 350$ states/eV and $\xi \cong   7.7$a.  The frequency of the
relaxational mode, $\omega_{\rm SF}$, is of order
$10-20$meV, quite small compared to its value, the Fermi energy or
bandwidth, $\sim$eV, in a Landau Fermi liquid.  Away from $\rm Q_i$, for $\rm
{\bf q} \xi \agt 1$, the dynamical susceptibility tends to be Fermi-liquid like,
taking the form,

\begin{equation}
\rm \chi_{FL}({\bf q},\omega) = {\chi_q \over 1-i\omega/\Gamma_q}
\end{equation}

\noindent where $\rm \Gamma_q$ is comparable to the bandwidth, and $\chi_{\bf
q} \sim \chi_0$.

NMR experiments \cite{6} on quantities dominated by $\rm \chi_{NAFL}$, the
$^{63}$Cu spin-lattice relaxation rate,
$\rm ^{63}T_1^{-1}
\sim (T/\omega_{SF})$, and the spin-echo decay time, $\rm T_{2G} \sim \xi^{-1}$,
show that $\rm \chi_{NAFL}({\bf q},\omega)$ varies dramatically with doping and
temperature through changes in
$\rm \chi_{Q_i}$, $\omega_{\rm SF}$, $\xi$, and the dependence of $\omega_{\rm
SF}$ on $\xi$. \cite{7}  Specifically, one finds that above a temperature, $\rm
T_{cr}$, $\chi_{\rm NAFL}$ displays mean field or RPA behavior, with
$\omega_{\rm SF}$, and $\xi^{-2}$, varying linearly with temperature,

\begin{displaymath}
\rm \omega_{SF}(T) \sim \xi^{-2}(T) \sim a+bT.
\end{displaymath}

\noindent Between $\rm T_{cr}$ and a second crossover temperature, $\rm
T_*$, $\chi$ displays $\rm z=1$ dynamical
scaling behavior, with

\begin{displaymath}
\rm \omega_{SF}(T) \sim \xi^{-1}(T) \sim c+dT
\end{displaymath}

\noindent The phase between $\rm T_{cr}$, and $\rm T_*$, is called
``pseudoscaling" because the scaling behavior found there is not universal.
Below $\rm T_*$, one enters the pseudogap phase, in which the antiferromagnetic
correlations become frozen, while
$\omega_{\rm SF}$, after reaching a minimum, increases rapidly as the
temperature is further decreased.  ``Pseudogap" denotes the quasiparticle
gap-like behavior found between $\rm T_*$ and $\rm T_c$, a behavior which is not
accompanied by the long range order of an antiferromagnet or a superconductor.

The two crossover temperatures, $\rm T_{cr}$, and $\rm T_*$, are, within
experimental error, the same as those found in an analysis of the uniform
magnetic susceptibility \cite{7}, which, unlike that of the familiar Landau
Fermi
liquid, is highly temperature dependent in all but highly overdoped systems.
Specifically, in both optimally doped and underdoped systems, the uniform
susceptibility at high temperatures increases with decreasing temperature,
displays a maximum at
$\rm T_{cr}$, following which it decreases linearly as the temperature is
further decreased, until the  second crossover temperature, $\rm T_*$, is
reached, below which it decreases more rapidly, down to $\rm T_c$.
Moreover, as shown in Fig. 1, the crossover
behavior found magnetically in probes of both long and short wave length
behavior is found as well in transport.

The phase diagram one obtains from an analysis of the NMR and INS experiments is
depicted in Fig. 2.  Note that from a magnetic perspective, optimally doped
systems are in fact underdoped; only when one enters the magnetically overdoped
regime at substantively higher doping concentrations, does one find an almost
temperature independent uniform susceptibility, and mean field behavior for
$\chi_{\rm NAFL}({\bf q},\omega)$ at all temperatures above
$\rm T_c$.  The detailed analysis of NMR and INS experiments presented in Refs.
(7) and (3) makes it possible to obtain a criterion for $\rm T_{cr}$ in terms
of the strength of the antiferromagnetic correlations $\xi$, at $\rm
T_{cr}$; one finds

\begin{equation}
\rm \xi(T_{cr}) \cong 2a
\end{equation}

\section*{\bf Transport Properties}

It is often stated that for optimally doped systems the longitudinal
resistivity, $\rm \rho_{xx}$, is linear in T in the normal state, while the
cotangent of the Hall angle, ctn $\rm \theta_H$, displays $\rm T^2$ behavior.
However, as discussed in some detail in Ref. (8), experiments using single
crystals show that $\rm \rho_{xx}(T)$ displays a downturn from linear in T
behavior as T
approaches $\rm T_*$, and that such downturns are a common feature of
all magnetically underdoped materials.  As may be seen in Fig. 3, the character
of the departure of $\rm \rho_{xx}$ from linear in T behavior depends on
doping level; optimally doped systems display the least departure from T-linear
behavior, while magnetically overdoped systems display an upturn at
comparatively high temperature.

In the magnetically underdoped cuprates, the transverse conductivity, $\rm
\sigma_{xy}$ is $\rm \sim T^{-3}$ at high temperatures, with significant
deviations occurring at temperatures $\alt T_*$.  For example, in
YBa$_2$Cu$_3$O$_{6.63}$, one finds $\rm \sigma_{xy} \sim T^{-4}$ below $200$K
($\rm \sim T_*$ for this material).  As a result, in the pseudoscaling regime,
the Hall resistivity, $\rm \rho_{xy} \cong (\sigma_{xy}/\sigma^2_{xx})$, is
temperature dependent ($\rm \sim T^{-1}$), while ctn$\rm \theta_H \equiv
(\rho_{xx}/\rho_{xy})$ displays $\rm T^2$ behavior.  Departures from this
behavior are found in the pseudogap regime and may occur as well in the
mean field regime.  For example, Hwang {\it et al.} \cite{9} find that in the
La$_{2-x}$Sr$_x$CuO$_4$ system, for $x \agt 0.15$, the Hall resistivity
becomes temperature dependent below a characteristic temperature $\rm \sim
T_{cr}$.  Thus the crossovers at $\rm T_{cr}$ and $\rm T_*$ seen in the spin
response possess direct counterparts in the charge response of the planar
quasiparticles in the cuprate superconductors.

\section*{\bf Quasiparticle Properties and Fermi Surface\\
 Evolution}

Recent specific heat \cite{10} and angle resolved photoemission (ARPES)
experiments \cite{11} show that in underdoped systems the quasiparticle
spectrum in the normal state is also anomalous.  A consistent account of
specific heat experiments can be obtained if one assumes that below $\rm
T_{cr}$ the energy spectrum of quasiparticles near the Fermi surface becomes
temperature dependent in such a way that the effective quasiparticle density of
states, $\rm N_0(T)$, mimics the temperature dependence found for $\rm
\chi_0(T)$ \cite{10}.  Still more detailed information concerning the
temperature dependent quasiparticle spectrum comes from ARPES experiments on
the BSCCO system, which show that a distinct evolution in the Fermi surface
takes place at temperatures $\alt 200$K in the underdoped systems, while for
overdoped BSCCO, where one finds a large hole Fermi surface consistent with
Luttinger's theorem, no such evolution is observed \cite{11}.  As Z. X. Shen has
told us at this meeting, in underdoped  quasiparticles on the Fermi surface
which are located in the vicinity of
$(\pi,0)$ become gapped, with a leading edge gap $\sim 20$meV and a spectral
function which has a broad maximum at $100-200$meV.

\section*{\bf The Nearly Antiferromagnetic Fermi Liquid (NAFL) Model}

How can a magnetic interaction between quasiparticles bring about this
remarkable normal state behavior?  In the NAFL description of the normal state
\cite{2}, the dominant contribution to the magnetic interaction between planar
quasiparticles is assumed to come from spin-fluctuation exchange, and so will
be proportional to $\chi$, where $\chi$, wherever possible, is taken from
experiment.  The quasiparticle spectrum is assumed to take a tight-binding form,

\begin{equation}
\varepsilon_k = -2t(cos k_x a + cos k_y y)-4t^{\prime} cos k_x a cos k_y
a-2t^{\prime\prime} [cos 2 k_x a + cos 2k_y a]
\end{equation}

\noindent where $t$, $t^{\prime}$, and $t^{\prime\prime}$ are the appropriate
hopping integrals, chosen to obtain agreement with ARPES experiments.  Thus the
effective magnetic interaction between quasiparticles for momentum transfers in
the vicinity of {\bf Q} is assumed to be

\begin{equation}
V^{NAFL}_{eff}({\bf q},\omega) = g^2_1\, \chi_{NAFL}({\bf q},\omega)
\end{equation}

\noindent and elsewhere to be

\begin{equation}
V^{FL}_{eff}({\bf q},\omega) =  \rm g^2_2\, \chi_{FL}({\bf q},\omega).
\end{equation}

\noindent Here $\rm \chi_{NAFL}$ is specified by Eq. (1) with parameters taken
from fits to INS and NMR experiments, while $\rm \chi_{FL}$ takes the general
form, Eq. (2), with the band parameters specified by Eq. (4).  The system
resembles a Fermi liquid in that it possesses a well-defined Fermi surface which
for doping levels near or beyond optimal satisfies Luttinger's theorem, and spin
and charge are not separated; both magnetic and transport properties derive from
the interaction, Eqs. (5) and (6).  However, because the interaction is very
strong and sharply peaked near $\rm {\bf Q}$, {\it none of the other
quasiparticle properties resemble those of a conventional Landau Fermi liquid};
hence the proposal that the system is, instead, a quite new kind of Fermi
liquid, with anomalous spin and charge properties which are doping and
temperature dependent.

An interesting feature of the effective interaction, $\rm V_{eff}^{NAFL}({\bf
q}\omega)$, is that it is temperature dependent; since $\rm \chi_{\bf Q}$
scales with $\xi^2$, in both the mean field and pseudoscaling regimes the
effective quasiparticle interaction becomes stronger as the temperature
decreases.  $V^{NAFL}_{eff}({\bf q},\omega)$ depends on the properties of the
quasiparticles through $\omega_{\rm SF}$ as well.  Moreover, as the interaction
becomes stronger, the behavior of the quasiparticles is modified substantially;
both quasiparticle energies and the Fermi surface can develop a substantial
temperature dependence.  One thus has a system in which, since the effective
interaction both modifies quasiparticle behavior and is itself altered by that
changed quasiparticle behavior, non-linear feedback, either negative or
positive, can play a significant role.

To the extent that one is able to determine both $\rm \chi^{NAFL}({\bf
q},\omega)$ and the quasiparticle spectrum from experiment (as is the case for
YBa$_2$Cu$_3$O$_7$), in calculations
of system properties based on the model interaction, Eq. (5) and the
quasiparticle spectrum, Eq. (4) one is left with only one free parameter.  One
test, then, of the correctness or utility of this proposed magnetic interaction,
is whether the resulting calculations yield agreement with experiment for a
number of system properties as one varies temperature and doping.  We shall see
that this is indeed the case:  starting from Eqs. (4) and (5), it is now
possible to explain the temperature and doping dependence of quasiparticle
lifetimes, the longitudinal resistivity, the Hall conductivity, $\rm
\sigma_{xy}$, the optical properties, the general features of Fermi surface
evolution in underdoped cuprates, and the transition at $\rm T_c$ to a $\rm
d_{x^2-y^2}$ pairing state.

A second test would be the derivation of the model interaction, Eq. (5),
from first principles.
Here one is, at present, somewhat less successful.  As discussed in Refs. (12)
and (13), one can, under some conditions, obtain a dynamic susceptibility of the
form, Eq. (5), from the effective 2D one-band Hubbard model; moreover one can,
at high temperature and for some doping levels, show that the effective
interaction obtained in a Hubbard-based Monte Carlo calculation is equivalent to
the model interaction specified by Eqs. (4) and (5) \cite{14}.  However, quite
generally, a major source of difficulty for microscopic calculations stems from
the fact that significant contributions to the real part of $\chi$ come from the
incoherent part of the quasiparticle spectral density, a region for which the
spectral density is imperfectly known.

\section*{\bf Hot and Cold Quasiparticles}

Because the effective interaction in a NAFL is, by definition, highly
ani\-so\-tro\-pic, 
the resulting quasiparticle behavior as one moves around the Fermi
surface necessarily reflects this anisotropy.  As shown in Fig. 4, there will,
in general, be two distinct groups of quasiparticles on the Fermi surface:  hot
quasiparticles are those located in the vicinity of hot spots, regions of the
Fermi surface where the magnetic interaction is determined by $\rm
V^{NAFL}_{eff}({\bf q},\omega)$ and is anomalously large.  Cold quasiparticles
are found in the remaining parts of the Fermi surface, where the interaction is
``normal," i.e. comparable to that found in conventional Landau liquids.  Both
the Eliashberg calculations of Monthoux and Pines
\cite{15} and the perturbation theoretic analytic calculations reported in
Stojkovic and Pines (SP) \cite{8} show that the lifetimes of hot and cold
quasiparticles differ significantly, as do their contributions to the transport
coefficients in the normal state.  Explicit
calculations, subsequently borne out by an analysis of experiments on
$\sigma_{xx}$ and $\sigma_{xy}$, show that in the vicinity of $\rm T_*$,
$\rm \tau^{-1}_{hot}$ crosses over from being nearly independent of T
(above $\rm
T_*$) to becoming linear in T, while over much of the temperature range of
interest, $\rm \tau^{-1}_{cold} \sim T^2$; it is this latter quantity which
is responsible for the measured temperature dependence of ctn $\theta_H$.
Stojkovic and I found that the changes with doping and temperature in $\rm
V^{NAFL}_{eff}({\bf q},\omega)$ combine with the
changes in the quasiparticle spectrum (measured in ARPES experiments) to produce
the measured crossovers in
$\sigma_{\rm xx}$, and $\sigma_{\rm xy}$, as one varies doping and temperature.
I refer the interested reader to Ref. (8) for a detailed account of the
resulting theory of the longitudinal and Hall conductivities, a theory which
appears capable of accounting for the rich morphology found experimentally in
the superconducting cuprates.

\section*{The NAFL Description of the Dynamic\\
 Susceptibility}

In the NAFL description of $\rm \chi({\bf q},\omega)$, since the interaction is
of electronic origin, the crossovers at $\rm T_{cr}$ and $\rm T_*$ reflect
changes in quasiparticle behavior; specifically, in the
behavior of quasiparticles located in hot regions of the Fermi surface.
Consider the behavior of $\chi({\bf Q}_i,\omega)$, the susceptibility at  one of
the (generally) incommensurate peaks.  On writing

\begin{equation}
\chi({\bf Q}_i,\omega) = {\chi_{{\bf Q}_i} \over 1-i\omega/\omega_{SF}} =
{\tilde{\chi}_{{\bf Q}_i} \over 1-J_{{\bf Q}_i}\tilde{\chi}_{{\bf Q}_i}(\omega)}
\end{equation}

\noindent where $\tilde{\chi}_{{\bf Q}_i}$ is the irreducible particle hole
susceptibility associated with quasiparticle transitions from one hot spot to
another, and $J_{{\bf Q}_i}$, the effective restoring force in the
particle-hole channel, may safely be assumed to be temperature independent, it
is evident that the measured temperature dependence of $\xi$ and
$\omega_{\rm SF}$ originate in the temperature dependence of $\tilde{\chi}_{{\bf
Q}_i}(T)$.  A closer examination shows that $\tilde{\chi}^{\prime}_{\bf Q}$ is
only weakly temperature-dependent; it is $\tilde{\chi}^{\prime\prime}({\bf
Q}_i,\omega)$, or, what is equivalent, the quasiparticle damping of spin motion,
which changes character dramatically as the temperature is decreased.  Thus, as
discussed in Monthoux and Pines \cite{16}, and in Chubukov, Pines and Stojkovic
(CPS) \cite{12}, on writing

\begin{equation}
\tilde{\chi}^{\prime\prime}({\bf Q}_i,\omega) \cong
\tilde{\chi}^{\prime}{({\bf Q}_i},\omega) N_{\bf Q_i}(T)\omega
\end{equation}

\noindent one sees that above $\rm T_{cr}$, in the mean field regime,
$\rm N_{\bf Q}(T) \sim 1/\tilde{\Gamma}_{\bf Q} \sim N_{\bf Q}$
($\Gamma_{\bf Q}$
is an effective band width), is essentially independent of T, while between
$\rm T_*$ and $\rm T_{cr}$, the damping becomes strongly T dependent in such a
way that one has
$\rm N_{\bf Q}(T) \sim 1/\xi(T) \sim a + bT$ so that $\omega_{SF} \sim 1/N_{\bf
Q}(T)\xi^2 \sim \hat{c}/\xi(T))$, corresponding to non-universal $\rm z=1$
pseudoscaling.  Below $\rm T_*$, in the pseudogap regime, $\xi \rightarrow$
constant, while one finds a more rapid fall off with decreasing temperature of
the quasiparticle damping, as seen in the ``effective" density of states
$\rm N_{\bf Q}(T)$; it is this which is responsible for the corresponding
``pseudogap behavior" seen in $\omega_{\rm SF}$.  The behavior of $N_{\bf Q}(T)$
is similar, \underline{but not identical}, to that one infers for the
quasiparticle state density, $N_T(0)$, from the measured behavior of
$\chi_0(T)$.

\section*{A Magnetic Scenario for Pseudogap Behavior}

Now that a near-consensus on the nature of the pairing state has been reached
\cite{17}, explaining pseudogap behavior, the sequence of crossovers seen in
the normal state of optimally doped and underdoped cuprate superconductors, is
arguably the major challenge facing the high-$\rm T_c$ community.  Two
scenarios for pseudogap behavior have been fleshed out in some detail:  a
d-wave superconductivity precursor scenario, in which a quasiparticle gap with
$\rm d_{x^2-y^2}$ symmetry appears above $\rm T_c$ due to superconducting
fluctuations \cite{18}, and a magnetic scenario, in which the hot quasiparticles
become gapped by the formation of a precursor to a spin density wave \cite{12}.
In the (pre-formed pair) SC scenario, the quasiparticle gap near $(0,\pi)$
observed in ARPES experiments is due to BCS pairing, while in the magnetic
scenario, it is produced by an SDW precursor which acts primarily on the hot
quasiparticles.  As noted in CPS, both scenarios imply that the quasiparticle
gap near $(0,\pi)$ should not change as the system becomes superconducting, in
agreement with the data.  There would, however, seem to be at least three
reasons for preferring the magnetic scenario:

$\bullet$ The doping dependence of $\rm
T_{cr}$, which marks the onset of pseudogap behavior (recall that for $\rm T <
T_{cr}$, $\chi_0$ starts to decrease with decreasing temperature) is markedly
different from that found for $\rm T_c$, posing a significant hurdle for a
preformed pair scenario.  On the other hand, in the magnetic scenario, there is
no reason for $\rm T_{cr}$ and $\rm T_c$ to be related.  Moreover, since
$\xi(\rm
T_{cr})$ has been shown to be $\sim 2$a, in both the 1-2-3 and 2-1-4 systems,
the doping dependence of $\rm T_{cr}$ is a natural consequence of the
increasing strength of the AF
correlations as one goes from overdoped systems toward the AF insulator.

$\bullet$ As noted by CPS, the magnetic scenario correctly describes the entire
sequence of crossovers in the normal state, including the crossover at $\rm
T_{cr}$ to $\rm z=1$ pseudoscaling, a crossover which is difficult to explain
as an SC precursor.

$\bullet$ It appears extremely difficult, if not impossible, to use a d-wave
precursor scenario to explain the onset of pseudogap behavior in
YBa$_2$Cu$_4$O$_8$ which Curro {\it et al.} \cite{5} find takes place at a
temperature, $\rm T_{cr}$, $\sim 6 \rm T_c$.

Let us examine the magnetic scenario in more detail.  At temperatures large
compared to $\rm T_{cr}$, strong coupling effects dominate; the scattering of
quasiparticles against each other reduces the uniform susceptibility, so as T
decreases (and the quasiparticle lifetime increases) $\chi_0$ will increase, an
effect seen in the Eliashberg calculations of Monthoux and Pines \cite{15}.
However, this tendency is opposed by the increasing strength of the AF
correlations, which may be expected to suppress $\chi_0$ (a quasiparticle whose
motion is strongly correlated with that of its neighbors is less able to
respond to an external magnetic field).  $\rm T_{cr}$, where $\xi \sim 2$a, is
then the temperature at which these two competing effects are roughly in
balance;
below $\rm T_{cr}$, the AF correlations dominate.  Put another way, when
$\xi \sim
2$a, in a configuration space description this means that a quasiparticle (the
hybridized Cu$^{2+}$ spin and hole) must now move in such a way that its motion
is correlated on average with some twenty-four of its neighbors.  Such motion
is best described as the precursor to a spin-density wave state.

In a momentum space description, as discussed in CPS, it is the hot spot
quasiparticles which are primarily affected by the spin-density wave precursor
formation; below $\rm T_{cr}$, the coherent part of the hot quasiparticle
Green's
function becomes progressively weakened, until at $T_*$, the end point of the
crossover, the system actually begins to lose pieces of the Fermi surface.  This
Fermi surface evolution is clearly seen in the ARPES experiments of the Stanford
and Argonne groups.  Its equivalent appears in the
$T=0$ studies of the NAFL by Chubukov and his collaborators \cite{12,19}
who find
two crossovers in an NAFL as one changes doping and the coupling between
quasiparticles increases, crossovers which it is natural to associate with those
at $\rm T_{cr}$ and $T_*$ observed at fixed doping and finite temperatures, as
discussed by CPS.  Very recent finite temperature NAFL calculations by Vladimir
Roubtsov show this onset of the Fermi surface evolution with temperature for
sufficiently strong coupling.

It seems natural, therefore, to associate the ``leading edge" gap found in ARPES
experiments with the influence of the precursor SDW on the``hot" quasiparticles,
and to designate that gap as $\Delta_{hot}$.  Although brought about by a near
approach to AF behavior (with $\xi$, in general, large compared to 2a), it
resembles a ``precursor" pairing gap, in that a substantial build-up
of a peak in $\rm \chi_{\bf Q}$ automatically produces strong ``pairing"
correlations between adjacent, barely itinerant, $\rm Cu^{2+}$ spins.  The
mathematical consequences of this pairing are, however, different from
conventional BCS pairing, since the fall-off in $\rm \chi_0(T)$ between $\rm
T_*$ and $\rm T_c$  is concave downwards, while that produced by BCS pairing
between $\rm T=0$ and $\rm T_c$ is concave upwards.  The pairing is
$\rm d_{x^2-y^2}$-like, since the SDW-induced gap is maximum for hot
quasiparticles, and is zero for cold quasiparticles maximally distant from the
hot regions.  Both $\Delta_{hot}$ and the strength of the AF correlations are
essentially fully formed at temperatures a little below $\rm T_*$, since
non-linear feedback effects associated with the appearance of $\rm \Delta_{hot}$
will prevent the real part of the static irreducible particle-hole
susceptibility from increasing appreciably beyond its value at $\rm T_*$ (recall
that $\rm \chi_{\bf Q} =
\tilde{\chi}_{\bf Q}/1-J_{\bf Q}\tilde{\chi}_{\bf Q}$, so  a freezing of
$\tilde{\chi}_{\bf Q}$ via $\rm \Delta_{hot}$ acts to freeze $\xi$).  Note that
between $\rm T_*$ and $\rm T_c$, while the hot quasiparticles
change their character, the cold quasiparticles are relatively unaffected until
$\rm T_c$ is reached.

On the above picture, the existence of two classes of quasiparticles in the
normal state, hot and cold, translates directly into the existence of two kinds
of quasiparticle energy gaps.  The first to appear, and largest in magnitude, is
$\rm \Delta_{hot}$, which may be related to $\rm T_*$, but which is almost
certainly \underline{not} related to $\rm T_c$.  Superconductivity comes about
only when one gets BCS pairing of the ``cold" quasiparticles; the pairing state
will definitely be $\rm d_{x^2-y^2}$, with a maximum gap, $\rm \Delta_{cold}$,
which is proportional to $\rm T_c$, and can easily be of order  $3-3.5 \rm
kT_c$.  The symmetry of $\rm \Delta_{cold}(T)$, as well as its temperature
dependence (it is expected to reach its maximum magnitude at $\rm T \sim
T_c/2$), means that below $\rm T_c/2$ all thermally excited quasiparticles are
sitting near the nodes of $\rm \Delta_{cold}$; hence gap values deduced from NMR
or $\rm \lambda(T)$ measurements refer to $\rm \Delta_{cold}$.

On this scenario, the superfluid density, $\rm \rho_s$, refers only to the
``cold" quasiparticles, so that it reflects not the total density of
quasiparticles, but only those quasiparticles which have not been ``gapped" by
the precursor to an SDW.  It will be smaller for low hole density
superconductors in part because these contain a higher percentage of hot
quasiparticles which cannot participate in the s.c. behavior, tied up as
they are
by $\rm \Delta_{hot}$.

Finally, this scenario explains the observations that
below $\rm T_c$, the position of the leading edge does not change (it can't,
being set already by SDW physics well above $\rm T_c$), while it does sharpen
(because quasiparticles away from the hot spot no longer scatter freely against
the spin fluctuations).  It also provides alternative explanations
for the appearance of a new high energy scale, $\alt 200$meV, in the underdoped
systems.  One is that it is the spin gap, a collective mode with the energy
$\leq$ 2J - required to flip the spin of one of the hot SDW-paired
quasi-particles; a second, suggested by Branko Stojkovic (private communication)
is that it reflects the appearance of new van Hove singularities associated with
the evolution of the Fermi surface.

Since the magnetic scenario predicts two distinct energy gaps for underdoped
systems, with the larger ($\rm \Delta_{hot}$) varying little with doping, the
smaller ($\rm \Delta_{cold}$) scaling with $\rm T_c$, a systematic
identification of these gaps and study of their doping dependence would provide
an important test of its applicability.

\section*{The Pairing Potential and Vertex Corrections}

A key question concerning the NAFL model is the role played by vertex
corrections to the strong coupling Eliashberg calculations of normal state
properties and the superconducting instability by Monthoux and Pines
\cite{15}.  Early estimates by Monthoux (private communication) showed that for
optimally-doped YBCO, where $\rm \xi(T_c) \sim 2a$, these would not be large
$(\alt 10-20\%)$, while Schrieffer \cite{20} has argued that for the underdoped
systems, where the correlation length is long, one is so close to SDW behavior
that vertex corrections will dramatically reduce the magnetic interaction
between quasiparticles and the pairing potential for the transition to the
superconducting state.  The NAFL model calculations at $\rm T=0$ by Chubukov
and his collaborators \cite{19,13} show that as one reduces the doping
concentration, and hence increases the coupling strength of the magnetic
interaction, vertex corrections initially act to increase this interaction and
hence enhance the pairing potential; however in the limit of very strong
coupling, where a preformed SDW has brought abou;substantial changes in the
quasiparticle Fermi surface, vertex corrections do bring about the substantial
reduction in the quasiparticle coupling to spin excitations proposed by
Schrieffer.  Quite recently, Monthoux \cite{21} has carried out a systematic
study at finite temperatures of the ``two-loop" corrections to his earlier
Eliashberg results.  He finds, in agreement with his earlier work, and with
Chubukov {\it et al.} that for systems near optimal doping, vertex corrections
are not large, and act to enhance the pairing potential, but that as the
magnetic correlation length increases, the magnitude of the vertex correction
to the effective interaction and pairing potential is quite different for hot
and cold quasiparticles; it is substantially larger for the hot quasiparticles,
and, for these, to first approximation, scales with $\xi$.

An interesting question, then, is whether in underdoped systems it is
Monthoux's vertex correction enhancement of the hot quasiparticle interaction
which brings about their transition to an SDW-paired state at temperatures $\rm
\sim T_*$, and whether, below $\rm T_*$, one then finds a hot quasiparticle gap,
$\rm \Delta_{hot}$, of the magnitude ($\sim 20$meV) required to explain the
ARPES experiments.  If so, one would have not only a ``proof of concept" for
the magnetic scenario for the underdoped systems, but would also be able to
understand the change in character of the vertex corrections as the temperature
decreases.  Thus for $\rm T \agt T_*$, vertex corrections enhance the magnetic
interaction between hot quasiparticles, while below $\rm T_*$, the coupling
of these ``SDW-gapped" quasiparticles to spin
excitations is indeed quite weak, being proportional to $\rm ({\bf Q}_i-{\bf
q})$, in agreement with the Ward identity arguments of Schrieffer.  On the
other hand, for optimally doped YBCO and for the higher $\rm T_c$ $\rm
T\ell$ and
Hg systems, where $\rm \xi(T_c) \sim 2a$, and, as well, for cold quasiparticles
in the 2-1-4 and other underdoped systems, vertex corrections will not play a
large role; Eliashberg calculations of the normal state properties, the pairing
potential and $\rm T_c$ should provide a quite reasonable approximation.

\section*{Acknowledgements}

I should like to thank Alexander Balatsky, Andrey Chubukov, Philippe Monthoux,
Doug Scalapino and Branko Stojkovic for stimulating discussions of the magnetic
scenario for underdoped systems.  This work was supported by an NSF Grant (NSF
DMR91-20000) to the Science and Technology Center for Superconductivity, and by
the Los Alamos National Laboratory.

\section*{Figure Captions}

Fig. (1)  A schematic depiction of the temperature crossovers measured for
various observable quantities in the underdoped cuprates.

Fig. (2)  A representative phase diagram for the cuprate superconductors.  Note
that the transition from underdoped to overdoped magnetic behavior occurs at a
concentration, $\rm x_{over}$, which is $\rm >x_{opt}$, the concentration for
which, within a given system, the superconducting transition temperature, $\rm
T_c$, is maximal.

Fig. (3)  The measured deviation from linear in $\rm T$ behavior of $\rm
\rho_{xx}(T)$ for underdoped, overdoped, and optimally doped superconducting
cuprates.  From top to bottom, results are shown for a representative overdoped
system, a $\rm T_c = 15$K sample of $T\ell 2201$, an optimally-doped system,
``YBa$_2$Cu$_3$O$_7$," and an underdoped system, YBa$_2$Cu$_3$O$_{6.63}$.  The
quantity plotted is [$\rm (\rho_{xx}(T)-\rho_0/\alpha T$] where $\rho_0$ and
$\alpha$ are obtained from a fit to the high temperature, linear in $\rm T$,
part of the resistivity [from Ref. (8)].

Fig. (4)  A model of a Fermi surface in cuprates (solid line) and the magnetic
Brillouin zone boundary (dashed line).  The intercept of the two lines marks
the center of the hot spots on the Fermi surface, regions near $(\pi,0)$ which,
because they can be connected by the wave vector $\rm{\bf Q}_i$, are most
strongly scattered into each other.

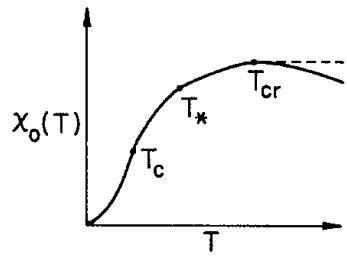
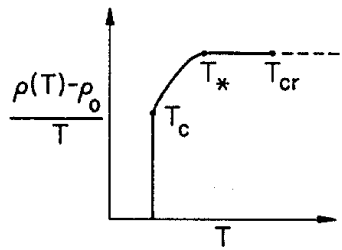
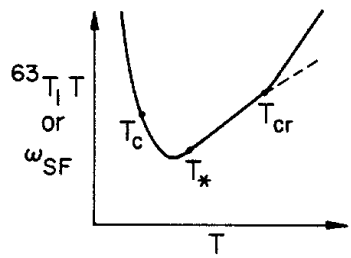
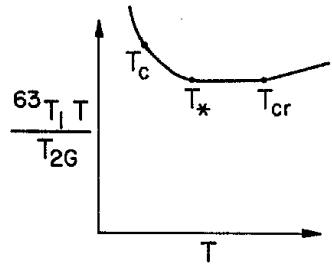
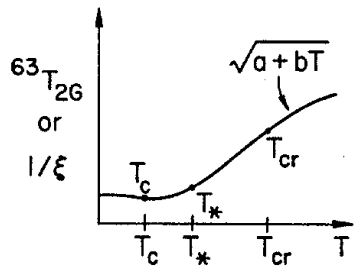
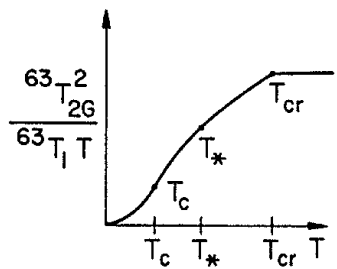
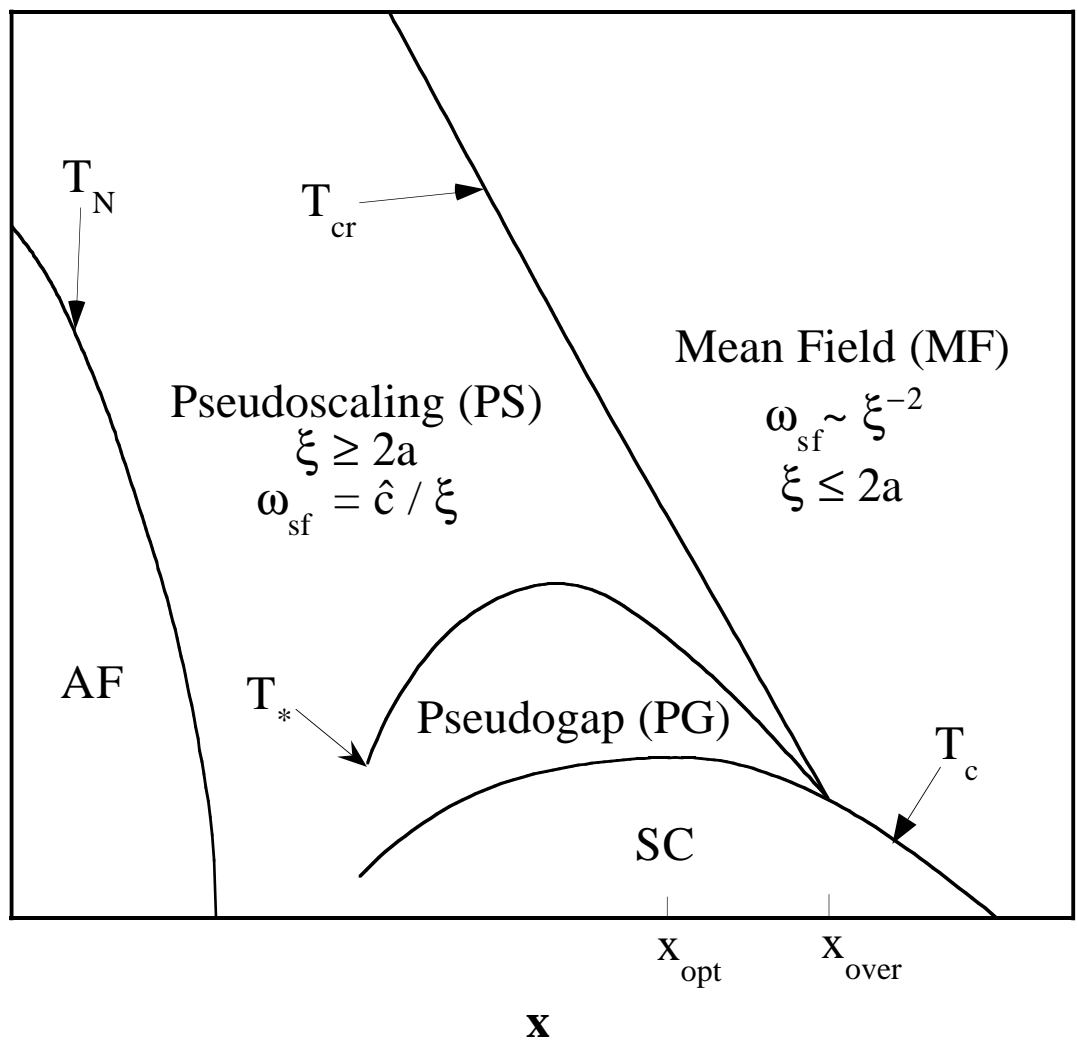
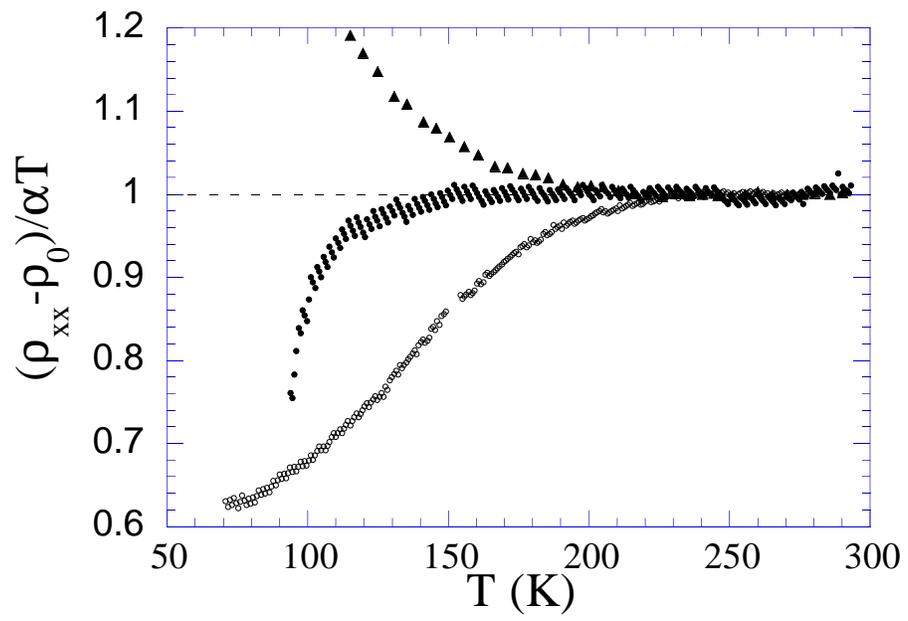
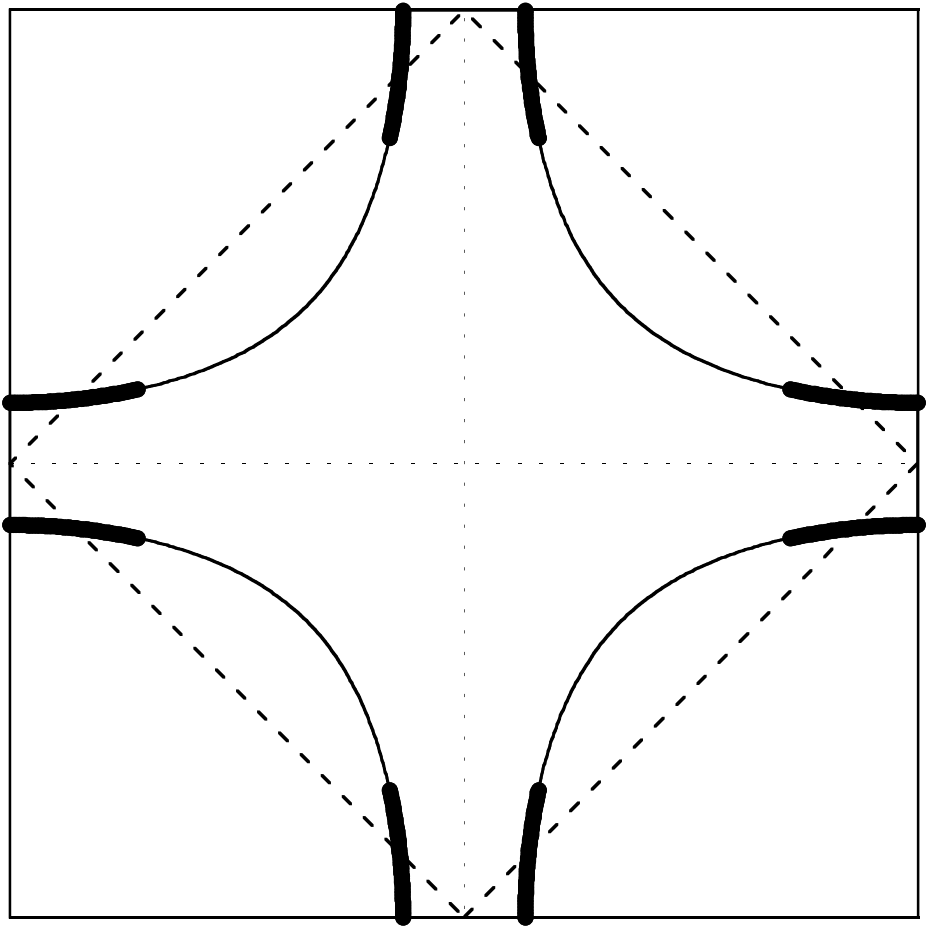
\end{document}